\documentclass[onecolumn,11pt]{article}
\usepackage{amsmath}
\usepackage{amsfonts}
\usepackage{amssymb}
\usepackage{amsthm}
\usepackage{times}
\usepackage{cancel}
\usepackage{graphics}
\usepackage{soul}
\usepackage{fixltx2e}
\usepackage{tikz}
\usepackage[normalem]{ulem}
\usepackage{mathrsfs}
\usepackage{lipsum} 
\usepackage{enumitem}
\usepackage[hang,flushmargin]{footmisc}
\setlist{nolistsep} 

\newcommand{\bpm}{\begin{pmatrix}}
\newcommand{\epm}{\end{pmatrix}}

\title{Quantum Logic and Quantum Reconstruction}
\author{Allen Stairs\\ \small Philosophy Department\\  \small University of Maryland, College Park, MD 20742, USA}
\date{}
\normalsize
\begin{document}

\maketitle

\begin{abstract}
Quantum logic understood as a reconstruction program had real successes and genuine limitations. This paper offers a synopsis of both and suggests a way of seeing quantum logic in a larger, still thriving context.

\bigskip
\noindent
\textbf{keywords:} quantum logic, quantum mechanics, lattice theory, orthomodular lattices, POVMs
\end{abstract}

\section{Introduction}
\label{intro}
The term ``quantum logic'' covers many philosophical views and more than one set of mathematical tools. However, lattices have been especially important in the history of quantum logic, and will be the focus of what follows. This paper doesn't aim to present new results, but rather to do a bit of summing up and looking forward. We begin with a brief review of the reconstruction of quantum mechanics---or Hilbert space, at least---within lattice theory. That will lead to a handful of small results with some foundational significance that can be demonstrated using elementary lattice theory. Then follows some discussion of limitations of lattices, a discussion of \textit{entwinement} and some associated puzzles, and finally some thoughts on broader approaches that are still within the spirit of quantum logic.

\section{Lattices and Representation Theorems}
\label{sec:2}
A lattice is a partially-ordered set $\mathcal{L} = \{L,\le\}$ with greatest lower bound and least upper bound operations $\land$ and $\lor$. If there are minimum non-zero elements, the lattice is atomic and those elements are atoms. $\mathcal{L}$ is complemented if there are least and greatest elements $\mathbf{0}$, and $\mathbf{1}$ and if for any $a$, there is at least one $b$ such that $a \land b = \mathbf{0}$ and $a \lor b = \mathbf{1}$. If there is a map $a \mapsto a^{\bot}$ such that $a$ and $a^\bot$ are complements, $a = a^{\bot\bot}$, and $a \le b \Rightarrow b^\bot \le a^\bot$ then $\mathcal{L}$ is orthocomplemented. An orthocomplemented lattice is orthomodular if $a \le b$ implies $b = a \lor(b \land a^\bot)$. We say that $a$ commutes with $b$ (write $aC b$) if $a = (a \land b) \lor (a \land b^\bot)$.  It can be shown that an orthocomplemented lattice is orthomodular if the relation $C$ is symmetric. 

Suppose $\mathcal{L}$ is an atomic orthomodular lattice. Suppose that in addition, $\mathcal{L}$ is complete (operations everywhere defined), irreducible (only $\mathbf{0}$ and $\mathbf{1}$ commute with every lattice element) and satisfies the covering law (if $p$ is an atom and $p \not\le q$, then there is no element strictly between $q$ and $p \lor q$). We'll call such lattices \textit{Piron Lattices}. Piron's 1964 result \cite{Piron1964} provides a representation theorem for such structures:

\begin{quote}
\textit{Piron's Theorem:} If $\mathcal{L}$ is a Piron lattice of dimension $\ge 4$, then there is an involutive division ring $D$ and a vector space $V$ with a Hermitian form (generalized inner product) $\langle\cdot,\cdot\rangle$ over $V$ such that $\mathcal{L}$ is isomorphic to the lattice $\mathcal{L}(V)$ of closed subspaces of $V$.
\end{quote}

If the goal of a reconstruction theorem is to get us all the way to Hilbert space, then Piron's theorem falls short. One obvious limitation: the theorem only holds in four or more dimensions. More importantly, in 1980 H. A. Keller showed that there are infinite-dimensional vector spaces with Hermitian forms over division rings other than $\mathbb{R}$, $\mathbb{C}$ and $\mathbb{H}$. \cite{Keller1980} In 1995, Maria Sol\`er identified the condition that picks out Hilbert space. Define orthogonality as usual: vectors $x$ and $y$ are orthogonal if $\langle x,y\rangle = 0$. For any set of vectors $M$, let $M^\bot$ be the set of all vectors orthogonal to every element of $M$.   $V$ is \textit{orthomodular}\footnote{Note that the term as used here is not the usual lattice-theoretic term.} if $S = S^{\bot\bot}$ for every closed subspace $S$. If $V$ also contains an infinite orthonormal sequence, then $V$ is a Hilbert space. \cite{Soler1995}. (Holland gives a clear exposition; see \cite{Holland1995}.)

Sol\`er's condition has a lattice-theoretic counterpart. Suppose that $\mathcal{L}(V)$ contains a sequence $\{a_1,a_2\ldots\}$ of orthogonal atoms such that for any pair $a_i,a_{i+1}$ there is an atom $c \le a_i \lor a_{i+}$ whose harmonic conjugate $c^\prime$ relative to $a_i,a_{i+1}$ is orthogonal to $c$. Then $V$ is a Hilbert space. (For  discussion and definitions, see Pitowksky \cite{Pitowsky2006} or Holland \cite{Holland1995}.) This condition entails that $c$ bisects the angle between the orthogonal atoms $a_i$ and $a_{i+1}$. (See \cite[p.12]{Pitowsky2006}.)

\section{Some Small Results}
\label{sec:3}

Quantum logic has obvious limitations as a reconstruction tool; more on that later. In this section, we look at a handful of small but hopefully interesting results that can be verified using only elementary lattice calculations.

\subsection{Entanglement}
\label{sec:3.1}

Schr\"odinger famously claimed that entanglement is \textit{the} characteristic feature of quantum systems. \cite[p. 555]{Schrodinger1935}. There's room to quibble about that, but any purported reconstruction of quantum mechanics has to address the question of how to represent compound systems. Whether or not a tensor product exists for every pair of Piron lattices, we can offer reasonable conditions on what such a product should be like when it does exist. The general idea is that the product should be a lattice of the same type, generated by independent isomorphic images of the subsystem lattices. I proposed the following conditions in \cite{Stairs1983}. Suppose that $\mathcal{L}_A$ and $\mathcal{L}_B$ are both Piron lattices.

\bigskip
\noindent
\emph{Product of Piron Lattices:}
\begin{enumerate}
\item $\mathfrak{L}_A\otimes\mathfrak{L}_B$ is a Piron lattice with least element $\mathbf{0}$ and greatest element $\mathbf{1}$
\item $i_A$ and $i_B$ map $\mathfrak{L}_A$ and $\mathfrak{L}_A$ isomorphically and independently\footnote{i.e., if $\alpha$ is any non-zero element of $\mathfrak{L}_A$ and $\beta$ is any non-zero element of $\mathfrak{L}_A$, then $i_A(\alpha)\land i_B(\beta)$ is a non-zero element of $\mathfrak{L}_A\otimes\mathfrak{L}_B$}  into $\mathfrak{L}_A\otimes\mathfrak{L}_B$, with $i_A(\mathbf{0}) = \mathbf{0}$, $i_A(\mathbf{1}) = \mathbf{1}$, and similarly for $i_B$.
\item If $\mathcal{B}_A$ and $\mathcal{B}_B$ are maximal Boolean sublattices of $\mathfrak{L}_A$ and $\mathfrak{L}_B$, then $i_A(\mathcal{B}_A)\cup i_B(\mathcal{B}_B)$ generates a Boolean product of $\mathcal{B}_A$ and $\mathcal{B}_B$ that is a maximal Boolean sublattice of $\mathfrak{L}_A\otimes\mathfrak{L}_B$
\item $i_A(\mathfrak{L}_A)\cup i_B(\mathfrak{L}_B)$ generates $\mathfrak{L}_A\otimes\mathfrak{L}_B$.
\end{enumerate}

\noindent
This definition generalize the requirements on product probability spaces. For the usual quantum cases, the product exists, though it is not unique. In particular, if $V_1$ and $V_2$ are Hilbert spaces, then $\mathcal{L}(V_1\otimes V_2)$ is a lattice product of $\mathcal{L}(V_1)$ and $\mathcal(V_2)$. The product conditions allow us to prove the existence of ``entangled'' atoms: atoms of $\mathfrak{L}_A\otimes\mathfrak{L}_B$ that can't be reduced to the form $i_A(\alpha)\land i_B(\beta)$, where $\alpha$ and $\beta$ are atoms of $\mathfrak{L}_A$ and $\mathfrak{L}_B$ respectively. For a detailed proof, see my \cite{Stairs1983}. Present purposes will be satisfied with a sketch. 

In what follows, we abuse notation and write $\alpha\land\beta$ rather than $i_A(\alpha)\land i_B(\beta)$, letting order keep track. Since our product generalizes the classical Boolean case, not every example will contain entangled elements. However, we can show that in the usual quantum case, entangled atoms must exist in any product. Let $\mathfrak{L}_A$ and $\mathfrak{L}_B$ be \textit{qubit lattices}---lattices of subspaces of two-dimensional Hilbert space. (Note that all such lattices are isomorphic, regardless of whether the vector spaces are over $\mathbb{R}$, $\mathbb{C}$ or $\mathbb{H}$.) Suppose $\{a,a^\bot\}$ and $\{b,b^\bot\}$ are pairs of orthogonal atoms in $\mathcal{L}_A$ and $\mathcal{L}_B$ respectively. Then a bit of lattice calculation shows that $a\land b$ and $a^\bot\land b^\bot$ are \textit{strongly perspective}. That is, $a\land b$ and $a^\bot\land b^\bot$ have a common complement in $\mathfrak{L}_A\otimes\mathfrak{L}_B$ (hence they are perspective) \textit{and} they have a common complement in their own span---within the sublattice whose zero is $\mathbf{0}$ and whose unit is $(a\land b) \lor (a^\bot\land b^\bot)$. Since $a\land b$ and $a^\bot\land b^\bot$ are atoms, this means that there is an atom $p \le (a \land b) \lor (a^\bot \land b^\bot)$ distinct from both $a \land b$ and $a^\bot \land b^\bot$---a ``superposition,'' if you will, of $a \land b$ and $a^\bot \land b^\bot$. Moreover, straightforward but tedious lattice calculation shows that this atom can't be identical with any atom of the form $i_A(\alpha)\land i_B(\beta)$. The argument can be generalized to higher dimensions. As we'll see, this has further consequences.

\subsection{No Cloning}
\label{sec:3.2}

Though hardly a complicated result, I'm not aware of this having been pointed out anywhere else. We can prove a lattice-theoretic version of the no-cloning result. Understood in lattice terms, a cloning machine would begin with a pair of isomorphic orthomodular lattices $\mathcal{L}_A$ and $\mathcal{L}_B$. If the initial state of the pair is represented by an atom $\phi_A\land\alpha_B$, the the cloning procedure would induce a change of state in accord with an automorphism $\iota$ on the lattice $\mathcal{L}_{A}\otimes\mathcal{L}_{B}$. The action of $\iota$ would be to map $\phi_A\land\beta_B$ to $\phi_A\land\phi_B$, where the automorphism doesn't depend on the choice of atom $\phi$to be cloned. (Here we assume that an isomorphism between $\mathcal{L}_{A}$ and $\mathcal{L}_{B}$ has been singled out, and that $\phi_B$ is the image of $\phi_A$ under this isomorphism.)

Suppose $\mathcal{L}_A$ and $\mathcal{L}_B$ are qubit lattices. Suppose that the state of the system to be cloned is represented by one of the atoms $a_1, a^\bot_1, a_2, a^\bot_2$, and the target system is in the ``ready state'' given by the atom $\beta$. In that case, $\iota$ will have to satisfy

\begin{enumerate}
\item $a_1\land \beta \mapsto_\iota a_1\land a_1$ 
\item $a^{\bot}_1\land \beta \mapsto_\iota a^{\bot}_1\land a^{\bot}_1$
\end{enumerate}

\noindent
Now suppose the state to be cloned is not $a_1$ or $a_1^\bot$, but $a_2$, which is an axis of perspectivity (``superposition'') of $a_1$ and $a_1^\bot$. Then $\iota(a_2 \land \beta)$ must be an axis of perspectivity of $\iota(a_1\land \beta)$ and $\iota(a_1^\bot\land\beta)$. But that amounts to saying that $a_2 \land a_2$ must be an axis of perspectivity for $a_1\land a_1$ and $a_1^\bot\land a_1^\bot$, and this is impossible. This is a special case of the point from \ref{sec:3.1}. The proof of the existence of entangled atoms amounts to showing that an axis of perspectivity for $a_1\land a_1$ and $a^\bot_1\land a^\bot_1$ can't have the form $\xi\land\psi$. If cloning is understood as a dynamical process represented by an automorphism on the lattice, then lattice-theoretic considerations rule out a general cloning procedure. Obviously this also shows that the measurement problem arises at a purely lattice-theoretic level.

\subsection{Why the Quantum? Lattices and PR Boxes}
\label{sec:3.3}

Quantum mechanics permits correlations that classical theories don't allow, but there are limits to their strength. Rather than saturating the CHSH inequality, quantum correlations can be no stronger than the Tsirelson bound allows. \cite{Tsirelson1980} On the other hand, there are logically possible systems that don't allow signaling but saturate CHSH. The most well-known examples are so-called ``PR boxes''---hypothetical non-local systems introduced into the literature by Popescu and Rohrlich. \cite{PopescuRohrlich1994}. The question ``Why the quantum?'' is often a way of asking for a principle that allows quantum correlations but rules out the ones that violate Tsirelson's bound. 

It would be a serious exaggeration to say that quantum logic has an answer to this question, but it does have something to say. (See \cite{StairsBub2013} for further discussion.) Once again, consider a pair of qubit lattices. We can ask: what limitations do the lattice conditions for $\mathcal{L}_A\otimes\mathcal{L}_B$ impose on correlations? Let's look at the particular case of PR box correlations and ask what would happen if we tried to model them by starting with two qubit lattices, reasoning about their product. Consider two systems, $A$ and $B$, with two binary observables on each. Let $\{a_1, a^\bot_1\}$ and $\{a_2, a^\bot_2\}$ represent the outcome pairs for $A$, and $\{b_1, b^\bot_1\}$ and $\{b_2, b^\bot_2\}$ play the same role for $B$. Intuitively, the PR box correlations are

\bigskip
\noindent
\begin{enumerate}
\item $a_1 \leftrightarrow b_1$
\item $a_1 \leftrightarrow b_2$
\item $a_2 \leftrightarrow b_1$
\item $a_2 \leftrightarrow b^{\bot}_2$
\end{enumerate}

\bigskip
\noindent
In the form of lattice polynomials, they become

\bigskip
\noindent
\begin{enumerate}
\item $(a_1 \land b_1) \lor (a^{\bot}_1 \land b^{\bot}_1)$
\item $(a_1 \land b_2) \lor (a^{\bot}_1 \land b^{\bot}_2)$
\item $(a_2 \land b_1) \lor (a^{\bot}_2 \land b^{\bot}_1)$
\item $(a_2 \land b^{\bot}_2) \lor (a^{\bot}_2 \land b_2)$
\end{enumerate}

\bigskip
\noindent
If 1---4 belonged to a common Boolean algebra, there would be no consistent way to assign the value $1$ to each; the Boolean conjunction of the four is $\mathbf{0}$, though any three are classically consistent. As a matter of pure lattice theory, it's also possible to represent any three of the four correlations; a Boolean algebra, after all, is a special case of a lattice. However, if the two possible $A$-inputs and the two possible $B$-inputs are each ``non-commuting,'' we are wrenched away from Booleanity; $a_2$ and $a^\bot_2$ are axes of perspectivity for $a_1$ and $a^\bot_2$ and $b_2$ and $b^\bot_2$ are axes of perspectivity for $b_1$ and $b^\bot_2$.

Reasoning from this stricture, the problem is easy to locate. It's not just that we can't represent all four correlations at once; it's that we can't even represent any pair. To illustrate, consider correlations $1$ and $2$. For the product lattice to embody both of these correlations, there would have to be an atom $\Phi$ in $\mathcal{L}_A\otimes\mathcal{L}_B$ such that $\Phi \le (a_1 \land b_1) \lor (a^{\bot}_1 \land b^{\bot}_1)$ and $\Phi \le (a_1 \land b_2) \lor (a^{\bot}_1 \land b^{\bot}_2)$. Because all elements of $1$ commute, as do all elements of $2$, we can re-write these two correlations as

\bigskip
\begin{enumerate}
\item $(a_1 \lor b^{\bot}_1) \land (a^{\bot}_1 \lor b_1)$
\item $(a_1 \lor b^{\bot}_2) \land (a^{\bot}_1 \lor b_2)$
\end{enumerate}

\bigskip
\noindent
For $\Phi$ to be under each of $1$ and $2$, it must be under each of $(a_1 \lor b^{\bot}_1)$, $(a^{\bot}_1 \lor b_1)$,
$(a_1 \lor b^{\bot}_2)$, and $(a^{\bot}_1 \lor b_2)$. Hence, $\Phi$ must be under the meet of any pair of these. But consider $(a_1 \lor b^{\bot}_1)$ and $(a_1 \lor b^{\bot}_2)$. $a_1$ commutes with each of $b^{\bot}_1$ and $b^{\bot}_2$. Therefore, $a_1$, $b^{\bot}_1$ and $b^{\bot}_2$ form a \textit{distributive triple}---all Boolean distribution rules apply. \cite[p. 466]{Holland1970}. We have 

\[
\Phi \le (a_1 \lor b^{\bot}_1) \land (a_1 \lor b^{\bot}_2) = a_1 \lor (b^{\bot}_1 \land b^{\bot}_2) 
\]

\noindent
Since $b^{\bot}_1$ and $b^{\bot}_2$ are atoms in  $\mathcal{L}_B$, their meet is $\mathbf{0}$. And so we have

\[
\Phi \le a_1.
\]

\noindent
But by an exactly similar argument using $(a^{\bot}_1 \lor b_1)$ and $(a^{\bot}_1 \lor b_2)$, we have

\[
\Phi \le a^\bot_1.
\]

\noindent
This would force $\Phi$ to be $\mathbf{0}$. However, $\Phi$ was introduced as an \textit{atom} in the support of both correlations; $\Phi$ cannot be $\mathbf{0}$. Thus, whatever the relationship between alternate choices of input may be for a PR box, it's of a different sort than the relationship between complementary quantum observables. 

\section{Limitations}
\label{sec:4}

A set of purely lattice-theoretic conditions picks out Hilbert space in the infinite-dimensional case. The proofs call for deep and difficult mathematics. The coordinatization theorems aren't adequate to pick out qubits (because the qubit lattice is only two-dimensional), and even with Sol\`er's result, we don't, so far as I know, have  an adequate characterization theorem for finite dimensions. Furthermore, even in the infinite dimensional case, Sol\`er's condition does not uniquely pick out the complex numbers. One might well be sympathetic to what Halvorson and Bub suggest: that quantum logic is a ``failed attempt'' at deriving quantum mechanics, and that to whatever extent it succeeds, it does so ``at the expense of complicating the axioms to the point of destroying all physical insight.'' \cite[p. 1]{HalvorsonBub2003}

Arguably, things aren't quite this dire. First a small point: the fact that Sol\`er's condition doesn't pick out $\mathbb{C}$ by itself isn't a problem. We can ask worthwhile questions about real and quaternionic quantum mechanics and so a general characterization theorem \textit{shouldn't} single out $\mathbb{C}$. This is perfectly compatible with looking for additional conditions that rule out $\mathbb{R}$ and $\mathbb{H}$.  More important, on what seems to me to be the most helpful understanding, the key idea of quantum logic is that the non-Boolean structure of events lies at the heart of the peculiarities of quantum mechanics. From that perspective, quantum logic is the study of the consequences of that structure---especially but not exclusively its consequences for probability. In this context, Itamar Pitowsky deserves special mention. It would be hard to read his work and come away with the impression that we get no physical insight from quantum logic. For example: Pitowsky pointed out that we can understand quantum uncertainty as a combinatorial, logical relationship (see, for example, \cite[section 2.1]{Pitowsky2003}). He also showed that Gleason's theorem \cite{Gleason1957} itself can be understood combinatorially. (For a sketch of the idea, see \cite[section 3.3]{Pitowsky2006}.) These are both non-trivial observations; the second is quite deep.

But what of the assumptions needed to establish Piron's and Sol\`er's theorems? Are they simply obscure? Most physicists don't speak the language of lattice theory, but nothing turns on that. Orthomodularity is a weakened form of the distributive law, which figures in innumerable standard probability calculations. If the lattice were fully distributive, it would be Boolean, and the point of quantum logic is that the events to which we assign probabilities don't have a Boolean structure. Further, orthomodularity is part of what ensures that Piron lattices are composed of Boolean sublattices ``pasted together,'' and this reflects another characteristic of quantum systems.  Non-commutativity of elements in an orthomodular lattice is the logical/lattice-theoretic counterpart of the non-commutativity of projectors in Hilbert space. The irreducibility condition is the lattice version of the fact that, setting superselection rules aside, the only projectors that commute with all projectors are $I$ and $\mathbf{0}$. Quantum logic doesn't ``explain'' noncommutativity; noncommutativity is inseparable from quantum theory. Every reconstruction builds it in one way or another. Quantum logic is distinctive in seeing noncommutativity as a feature of the logical structure of events themselves. 

Atomicity can't be justified \textit{a priori}, but it's hardly insight-destroying to add it to the list of conditions. There is a possible world with no least (i.e., maximally informative) non-zero experimental propositions; as far as we know, ours is not that world. 

The meaning of the covering law is less obvious, but as Jauch and Piron pointed out many years ago, it's equivalent to a plausible constraint on ideal measurement---that ideal measurements should preserve as much information as possible, consistent with the event structure.  (\cite{JauchPiron1969}. See also Wilce's discussion of quantum logic in the \textit{Stanford Encyclopedia of Philosophy}. \cite{WilceSEP}.)

What of Sol\`er's condition? Pitowsky hoped it could be understood as a probability interpolation requirement: if $a$ and $b$ are orthogonal atoms with $p(a) = 1$ and hence $p(b) = 0$, then there is a third atom $c \le a \lor b$ whose probability must be $1/2$. However, as Pitowsky makes clear \cite[section 3.4]{Pitowsky2006}, there are delicate issues here, including unsolved mathematical problems about finite dimensional cases.

\section{Entwinement and a Puzzle About POVMs}
\label{sec:5}

Schr\"odinger's comment notwithstanding, there's another feature of quantum theory that has at least as much right as entanglement to be considered the source of its mysteries, and it plays a central role in quantum logic. The word I would like to persuade people to use is \textit{entwinement}. In the prologue to proving his theorem, Gleason points out a difference between the frames (essentially, orthonormal bases) that we find in two dimensions and frames in all higher-dimensional cases. First we recall the term ``frame function.'' A frame function is an additive function $f$ on the frames of a Hilbert space such that there is a real number $r$ for which $\Sigma_i f\alpha_i = r$ for every frame $\{\alpha_i\}$. If $f$ is non-negative and $r = 1$, $f$ is a probability function. Gleason writes

\begin{quote}
In dimension two a frame function can be defined arbitrarily on a closed quadrant of the unit circle in the real case, and similarly in the complex case. In higher dimensions the orthonormal sets are intertwined and there is more to be said. (\cite[p.~886]{Gleason1957})
\end{quote}

We'll modify Gleason's term ever so slightly and say that two frames are \textit{entwined} if some but not all of their elements differ at most by a phase factor. We'll apply the same term to complete orthogonal sets of rays, to maximal Boolean sublattices that share some but not all of their atoms, and so on. The quantum-logical version of entwinement is shared atoms among maximal Boolean sublattices of $\mathcal{L}(V)$.

I would suggest that when people say that quantum mechanics is contextual, what they really mean is that exhibits entwinement. The point is that \textit{quantum theory} is \textit{not} contextual. Quantum mechanics represents entwined elements as being identical across contexts. The ``contextuality'' is only in the non-quantum constructions that would be called for to simulate the quantum behavior. Entanglement phenomena are a special case of entwinement. Howard et al. have recently argued that entwinement (contextuality, in their terms) lies behind the speed-up in quantum computation. \cite{HowardEtAl2014}. Their paper builds on work by Cabello et al. linking graph theory with entwinement. \cite{CabelloEtAl2014}

Gleason's theorem tells us that the probability structure of quantum theory is contained in the logical structure. Given the detailed entwinement relations among subspaces of a Hilbert space with three or more dimensions, the possible probability measures are completely determined, and in particular, assigning probability one to a one-dimensional subspace fixes all other probabilities. In two dimensions, however, there is no entwinement; probabilities on different frames are independent. 

This is related to an apparent shortcoming of quantum logic: it provides no account of generalized measurements, represented by positive operator-valued measures (POVMs). A POVM is a set $\{E_i\}$ of positive self-adjoint operators called \textit{effects}, such that $\Sigma_i E_i = I$. Surprisingly, there is a Gleason-style theorem for POVMs on two dimensional Hilbert space, as Busch showed in 2003. \cite{Busch2003}. Although a self-adjoint operator on two-dimensional Hilbert space has only two projectors in its spectral decomposition, POVMs on the same space can contain arbitrarily many effects. This extra structure is what makes the proof possible. Caves et al. \cite{CavesFuchsManneRenes2004} prove a collection of related results in which they explicitly use Gleason's ``frame function'' terminology. In this context, a frame function from effects  into $[0,1]$ that sums to $1$ for the effects in each POVM. Like the frames in Gleason's proof, POVMs are entwined with one another. Here, however, entwinement holds even for qubits. This means that combinatoric considerations fix the possible probability assignments for qubits as well as higher-dimensional systems. 

A related argument by Cabello \cite{Cabello2003} provides a Kochen-Specker-style no hidden variables proof and it will be worth reviewing that argument briefly. Let $A$, $B$ and $C$ be three one-dimensional projectors on two-dimensional Hilbert space, projecting onto three mutually non-orthogonal vectors. Consider the following three POVMs:

\bigskip
\begin{enumerate}
\item $\{\frac{1}{2} A, \frac{1}{2} A^\bot, \frac{1}{2} B, \frac{1}{2} B^\bot\}$
\item $\{\frac{1}{2} B, \frac{1}{2} B^\bot, \frac{1}{2} C, \frac{1}{2} C^\bot\}$
\item $\{\frac{1}{2} C, \frac{1}{2} C^\bot, \frac{1}{2} A, \frac{1}{2} A^\bot\}$
\end{enumerate}

\bigskip
\noindent
The effects in each set clearly add to $I$, and so each set is a POVM. Suppose that we think of the constituent effects for each POVM as representing mutually exclusive, jointly exhaustive outcomes and we add the assumption that the POVMs represent univocal quantities with \textit{noncontextual} predetermined outcomes. Then we are immediately led to a contradiction. Suppose, for example, that POVM 1 takes the value associated with $\frac{1}{2} A$. Then so must POVM 3. Furthermore, the system doesn't have any of the values associated with the $B$ and $C$ effects. But now POVM 2 can't take a value. It's clear that the same problem will arise no matter which effect we begin with. This appears to be a serious shortcoming of lattice-based quantum logic. Restricting our attention to subspaces misses structure that yields important results. 

POVMs have become increasingly important in discussions of quantum foundations. They give rise to structures with some features in common with lattice-based quantum logic, but with important differences. For a review of formal aspects of POVMs, see \cite{CattaneoEtAl2009}. Busch has contributed extensively to the development of the theory of POVMs. For a recent discussion of how POVMs might be incorporated into a realist interpretation of quantum theory, see \cite{BuschJaeger2010}. To be sure, there's more to a qubit than its lattice structure reveals. POVMs are widely seen as more fundamental than projection-valued measurements.  This may be true, but before we embrace the proposal, it's worth noting some puzzles. I'll describe them briefly; I discuss them in more detail in \cite{Stairs2007}. 

From one point of view, Cabello's argument can be taken as reinforcing an idea about POVMs that has been around for some time: they represent ``unsharp values.'' The example we've reviewed can be taken as a \textit{reductio} of the claim that POVMs \textit{could} represent sharp values. To see one problem, however, return to the three POVMs numbered 1---3 above. It's true that there's no way to select exactly one effect from each POVM. However, compare 1---3 with an analogous case for three qubits. Associate the $A$-effect with qubit one, the $B$-effect with qubit two, and the $C$-effects with qubit three. Consider these sets:

\bigskip
\noindent
\begin{enumerate}
\setcounter{enumi}{3}
\item $\{\frac{1}{2} A\otimes I\otimes I, \frac{1}{2} A^\bot\otimes I\otimes I, \frac{1}{2} I\otimes B\otimes I, \frac{1}{2} I\otimes B^\bot\otimes I \}$
\item $\{\frac{1}{2} I\otimes B\otimes I, \frac{1}{2} I\otimes B^\bot\otimes I, \frac{1}{2} I\otimes I\otimes C, \frac{1}{2} I\otimes I\otimes C^\bot \}$
\item $\{\frac{1}{2} I\otimes I\otimes C, \frac{1}{2} I\otimes I\otimes C^\bot, \frac{1}{2} A\otimes I\otimes I, \frac{1}{2} A^\bot\otimes I\otimes I \}$
\end{enumerate}

\bigskip
\noindent
It's easy to see that each operator is positive and each set adds to $I\otimes I\otimes I$. Thus, each set is a POVM on the eight-dimensional Hilbert space of the three qubits. But all six of the operators commute with one another. On the face of it, this would make it puzzling if the corresponding experimental propositions were intrinsically unsharp. It's still true: there's no way to select exactly one effect from each set. But what exactly this means is much less clear. In particular, there's nothing in the math to serve as a counterpart to the idea that, for example, $\frac{1}{2} A\otimes I\otimes I$ and $\frac{1}{2} I\otimes B\otimes I$ represent mutually exclusive outcomes. 

Even in the single qubit case, where the $A$, $B$ and $C$ operators are mutually non-commuting, it's not clear that $\frac{1}{2} A$, $\frac{1}{2} B$, etc. represent univocal experimental quantities. This means it's not clear that the ``entwinement'' is anything other than notional. For example, consider these experiments: (i) measure a qubit for $z$-spin and then for spin at $\pi/6$ radians to the $z$-axis in the $z-x$ plane; (ii) measure a qubit for $z$-spin and then for spin at $-\pi/6$ radians in the $z$-$x$ plane. These two experiments are distinct, and no matter what their outcomes, the qubit will be left in a different state in experiment (i) than in experiment (ii). However, the POVMs for the two experiments are identical. Sameness of effect \textit{qua} positive operator does not mean sameness of outcome, event, experimental proposition or resulting state. Experimental outcomes that clearly are \textit{not} physically equivalent are sometimes associated with one and the same effect. 

The point is certainly not that Busch or Cabello have made some sort of mathematical mistake, nor is the point that POVMs are in some way illegitimate. In particular, POVMs as probability bookkeeping devices are beyond reproach. Rather, we have an intriguing puzzle here. We have a rich structure with a clear kinship to quantum logic. It leads to consequences that apparently go beyond what standard orthomodular lattice-based quantum logic can account for. But there are questions about how best to understand the structure and the objects---effects and POVMs---from which it is constructed. Furthermore, some of these questions clearly raise logical issues: what are the criteria for settling when effect operators do and do not represent equivocally? How could a set of effects that apparently can't have sharp values be isomorphic to a set of mutually commuting effects? How should we understand attributions of unsharp values?

\section{Concluding Thoughts}
\label{sec:6}

I'll confess that I find the quantum logical reconstruction results---that is, the coordinatization theorems---remarkable. Discovering that purely lattice-theoretic assumptions can yield structures as rich as Hilbert space is a bit like one's first experience of seeing a rabbit pulled out of a hat. To be sure, the quantum logical reconstructions don't have the simplicity and elegance of Hardy's five-axiom reconstruction \cite{Hardy2001}. This is not really surprising. The quantum logical reconstruction aims to derive Hilbert space from a more basic level of structure than Hardy begins with; the assumptions used in the quantum logical reconstruction don't even mention probability. In spite of this, it's fair to say that a purely lattice-theoretic approach to quantum foundations is a program that has mostly run its course. However, this is entirely compatible with saying that we can learn a good deal about quantum systems by studying broadly logical and combinatoric features of quantum systems. \textit{That} program has by no means run out of steam. Pitowsky's work certainly fit that description. So does the work of Maria Dalla Chiara and her students. See \cite{DallaChiaraGiuntini2008}, \cite{CattaneoEtAl2009}. So does the work by Busch, Cabello and Caves et al noted above. So does the program of Cabello and his associates on the ``exclusion principle'' and related matters. See \cite{CabelloEtAl2013} and \cite{Cabello2013} in addition to the work already cited. Jeffrey Bub argues that the idea of ``quantum information'' is closely tied to the non-Boolean event structure of quantum mechanics. See, for instance, \cite{Bub2011}. There are many, many other examples. Much of this work doesn't draw on lattice theory. All of it, in my view, is ver much in the spirit of quantum logic.

There is no royal road to understanding quantum theory and it would be foolish to argue that broadly quantum logical approaches have some unique claim to insight. However, when the talk on which this paper is based was originally presented, the moderator suggested that I was about to discuss a ``failed program.'' With all due respect to that worthy gentlemen, I'm afraid I would have to disagree. It might be argued that the reconstruction theorems themselves didn't lead directly to physical insight, though that claim itself could be disputed. Those theorems did show, however, that a great deal of the physically important structure of quantum theory is implicit at what I have been calling the logical level. And there is an even better case for claiming that the broader concerns behind the interest in quantum logic have born real fruit. Whether or not quantum logic itself is dead, I expect its successors to continue making contributions for some time to come.




%

%

%


\bibliographystyle{unsrt}
\bibliography{quantumlogicbib.bib}

\end{document}